# On the in-situ detectability of Europa's water vapour plumes from a flyby mission

**Corresponding author**
Hans L.F. Huybrighs [a b c d]
huybrighs@mps.mpg.de
+49-551-384-979-527

Yoshifumi Futaana [a]
Stanislav Barabash [a]
Martin Wieser [a]
Peter Wurz [e]
Norbert Krupp [b]
Karl-Heinz Glassmeier [b, c]
Bert Vermeersen [d]

[a] *Swedish Institute of Space Physics, Post Office Box 812, Kiruna, Sweden*
[b] *Max-Planck-Institut für Sonnensystemforschung, Göttingen, Germany*
[c] *Institut für Geophysik und extraterrestrische Physik, Technische Universität Braunschweig, Mendelssohnstr. 3, Braunschweig, Germany*
[d] *Faculty of Aerospace Engineering, Delft University of Technology, Delft, Netherlands*
[e] *Physikalisches Institut, University of Bern, Sidlerstrasse 5, Bern, Switzerland*

## HIGHLIGHTS
- The in-situ detection of $H_2O$ and $H_2O^+$ from Europa's plumes is possible with JUICE/PEP
- This work focusses on plumes with a low mass flux (1 kg/s)
- The geometry of the plume source is not a major factor for the detectability
- Knowledge of Europa's exosphere is required to separate exospheric and plume $H_2O/H_2O^+$

# ABSTRACT

We investigate the feasibility of detecting water molecules ($H_2O$) and water ions ($H_2O^+$) from the Europa plumes from a flyby mission. A Monte Carlo particle tracing method is used to simulate the trajectories of neutral particles under the influence of Europa's gravity field and ionized particles under the influence of Jupiter's magnetic field and the convectional electric field. As an example mission case we investigate the detection of neutral and ionized molecules using the Particle Environment Package (PEP), which is part of the scientific payload of the future JUpiter ICy moon Explorer mission (JUICE). We consider plumes that have a mass flux that is three orders of magnitude lower than what has been inferred from recent Hubble observations (*Roth et al.*, 2014a). We demonstrate that the in-situ detection of $H_2O$ and $H_2O^+$ from these low mass flux plumes is possible by the instruments with large margins with respect to background and instrument noise. The signal to noise ratio for neutrals is up to ~5700 and ~ 33 for ions. We also show that the geometry of the plume source, either a point source or 1000 km-long crack, does not influence the density distributions, and thus, their detectability. Furthermore, we discuss how to separate the plume-originating $H_2O$ and $H_2O^+$ from exospheric $H_2O$ and $H_2O^+$. The separation depends strongly on knowledge of the density distribution of Europa's exosphere.

# 1   INTRODUCTION

Jupiter's moon Europa is thought to harbour a subsurface ocean of salty liquid water that could potentially support life (*Hand et al.*, 2009). The release of material from below the surface of Europa, water molecules for example, has been hypothesized to explain geological features on the surface (*Hoppa et al.,* 1999; *Fagents et al.,* 2000; *Nimmo et al., 2007; Quick et al., 2013*). However, past optical observations from Galileo and New Horizons have failed to prove the existence of any such releases (*Philips et al.*, 2000; *Hurford et al.,* 2007; *Roth et al.,* 2014a). On the other hand, during the E12 Galileo flyby of Europa a significantly raised electron density (*Kurth et al.*, 2001) was observed together with anomalously strong magnetic fields (*Kivelson et al.*, 2009). These atypical plasma conditions might be linked to active processes at Europa's surface, although they could possibly be explained the passing of a cold, dense blob of iogenic plasma as well (*Bagenal et al.*, 2015).

Hubble Space Telescope observations in December 2012 of ultraviolet emission lines of oxygen (at the 130.4 nm and 135.6 nm emission line) and hydrogen (Lyman-α) at Jupiter's moon Europa were interpreted as the existence of water vapour plumes near the south pole (*Roth et al.*, 2014a). Plume activity persisted during the 7 hours of observation time. The height of the plume derived from these observations is ~200 km. Furthermore the observations indicate a mass flux of 7000 kg/s and source kinetic temperature of the gas of 230K.

During similar Hubble observations in November 2012 and 1999 no plumes were observed. If a plume was present during these observations the density would have been two to three times lower (*Roth et al., 2*014a). Results were ambiguous for Hubble observations made in June 2008 (*Saur et al.,* 2011). Also, during repetitions of the successful observation in January and February 2014 (*Roth et al*., 2014b) and in the period between November 2014 until April 2015 (*Roth et al*., 2016) no plume signatures were detected. Such observations could imply that the plumes are not persistent or that the successful detection from December 2012 corresponded to an exceptionally strong event.

Nevertheless, the successful plume observation in December 2012 (*Roth et al., 2*014a) implies the existence of a localized source of neutral water molecules. The presence of the neutrals will result in the production of ions, mainly via electron impact ionization. The plume-originating neutrals and ions might be a potential source of the hypothetical Europa torus (*Lagg* 2003; *Mauk et al.,* 2003; *Mauk et al.,* 2004) or plasma plumes (*Intriligator et al.*,1982; *Russell et al.,* 1998 and *Eviatar et al.,* 2005) in addition to the exospheric particles sublimated or sputtered from the surface.

At Enceladus, one of Saturn's moons, ongoing plume activity is taking place with a global production of about 93 kg/s (*Waite et al*., 2006). In-situ plume sampling indicates the plumes are linked to Enceladus' subsurface ocean (*Postberg et al*., 2009; *Postberg* et al., 2011) and has allowed the study of Enceladus' subsurface ocean (see for example *Bouquet et al*., 2015). This suggests that if Europa's plumes are connected to its interior, sampling of the plume gasses could allow the study of the subsurface ocean. Some predictions have been made about the spreading of dust (*Southworth et al*., 2015) and bacteria-sized particles by Europa plumes (*Lorenz* 2015).

In this work we investigate the feasibility of in-situ measurements of Europa's plumes, by modelling the trajectories of neutral and ionized plume particles and the respective measurements by neutral and ion mass spectrometers. First, we use a Monte Carlo particle tracing method (also called test-particle method) to model the trajectories. Then, we simulate the first planned Europa flyby of the JUICE spacecraft that will take place on the 13th of February 2031. For this flyby we simulate the measurements conducted by the instruments for ion and neutral detection, and determine if they can detect the particles originating from the plume. We show that these instruments can provide a sufficiently high signal-to-noise ratio for plume-originating particles, even if we assume a plume that is three orders of magnitude less dense than the one reported in *Roth et al.* (2014a).

## 2 REFERENCE MISSION: JUICE

As a reference mission case we investigate the feasibility of plume particle detections by the Particle Environment Package (PEP) on-board of the ESA mission JUpiter ICy moon Explorer (JUICE). JUICE is scheduled to be launched in 2022 and will make two flybys of Europa in early 2031 (*Grasset et al.*, 2013). During the flybys JUICE will approach Europa up to a height of ~400 km.

The PEP experiment is composed of six particle instruments that will study the particle environment at Jupiter. In this study we simulate the observations of plume particles by the Jovian plasma Dynamics and Composition analyser (JDC) and the Neutral gas and Ions Mass spectrometer (NIM). JDC is a time of flight ion mass spectrometer that covers the energy range 1 eV – 41 keV; it can achieve a mass resolution (M/ΔM) up to 30. It also gives the energy and directional distribution of ions. NIM is a high mass resolution (M/ΔM > 1100) time of flight mass spectrometer, for low energy neutral and ionised particles (< 10 eV), based on an earlier version for lunar research (*Wurz et al.*, 2012). An overview of NIM and JDC features relevant to this work are shown in Table 1.

Table 1: JDC and NIM specifications

| JDC | |
|---|---|
| G, geometric factor per bin [cm$^2$ sr eV/eV] | 5.58×10$^{-4}$ |
| Efficiency ε | 0.035 |
| Maximum number of energy bins | 96 |
| Number of azimuth bins | 16 |
| Maximum number of elevation bins | 12 |
| Scan time per energy step [ms] | 7.8 |
| Full 3D scan (96×12×7.8ms) [s] | 9 |
| Energy range | 1 eV to 41 keV |
| Azimuth range [deg] | 360 |
| Elevation range [deg] | 90 |
| Expected JDC noise rate at Europa [counts/s] | 100 |
| | |

| **NIM** | |
|---|---|
| Conversion factor from count rate to density | 1 particle/cm$^3$ = 5 counts |
| Expected background counts at Europa (instrumental plus penetrating radiation) | 35 counts, for 5 seconds accumulation per time of flight (TOF) bin |
| Energy range | < ~5 eV |

# 3 NEUTRAL PLUME MODEL

We employed a Monte Carlo particle tracing method to model the neutral environment due to the Europa plume. In our model neutral plume particles are represented by super particles, a super particle (sometimes also called macro or meta particle) is a virtual particle that represents many (water) molecules travelling together (see for example *Holmström* 2006). The trajectories of the super particles are calculated under the influence of Europa's gravity by numerical integration. Particle collisions are not taken into account in our model, in the discussion section this assumption will be investigated in more detail. Non-collisional Monte Carlo particle tracing approaches have been used to model plumes on Enceladus (*Burger et al.*, 2007; *Smith et al.*, 2010; *Tenishev et al.*, 2010; *Dong et al.*, 2011, *Hurley et al.*, 2015) and Io (*Glaze and Baloga* 2000; *Doute et al.*, 2002). The model parameters used in our plume modelling are summarized in Table 2 and will be discussed in the following paragraphs.

Table 2: Model parameters for neutral plume model.

| Parameter | Value |
|---|---|
| Source temperature [K] | 230 |
| Source mass flux [kg/s] | 1 |
| Particles bulk speed (perpendicular to local surface) [m/s] | 460 |
| Mass per particle (mass H$_2$O molecule) [kg] | 2.987×10$^{-26}$ |
| Number of super particles for which trajectories are calculated | 8×10$^6$ |
| Time step length [s] | 10 |
| Number of time steps | 1000 |
| Size of a grid cell [km$^3$] | 10×10×10 |
| Mass per super particle [kg] | 8.75×10$^{-7}$ |
| Density in a super particle with the size of a grid cell (lowest density possible in the model) [#/cm$^3$] | 29 |

From the trajectories of a large set of super particles ($8\times10^6$), the spatial profile of the water molecule's number density is calculated. The number density is represented on an evenly separated grid. The grid size is 10 x 10 x 10 km. A time step of 10 seconds is used for the calculation of neutral particle trajectories, such that it is possible to capture a particle with a velocity resolution of 0.5 km/s (approximately the assumed bulk speed) within the same or neighbouring grid cell after a single time step. Then each particle will contribute to the density of every grid cell it passes through, without missing any cells. The reference frame used in this work is the IAU Europa centred reference frame (*Archinal et al.*, 2010). The *z*-axis is perpendicular to the mean orbital plane of Europa, the positive *x*-axis is pointing in the direction of Jupiter and the *y*-axis completes the right hand system (it is opposite to the sense of rotation of Europa about Jupiter and the direction the corotational plasma is flowing in).

We conducted simulations with two different source distribution models (shown in Figure 1). For the first case, the super particles are launched from a single point. For the second case, we assume a series of 200 point sources to simulate the release of water vapour from a 1000 km crack on Europa's surface. Such a crack corresponds to the Europa surface features typically referred to as 'Lineae' (see for example *Doggett et al.*, 2009). In the observations reported by *Roth et al.* (2014a) the source location and geometry cannot be identified because of the low spatial resolution of the images. However, they argue that the observed plume could originate from cracks on Europa's surface rather than a point source. It can be expected that the spatial density distribution will be different between the two cases, even when the mass flux between the two is the same, in particular close to the source. We assume a total mass flux both at the plume point source and crack of 1 kg/s. This is three orders of magnitude smaller than what was suggested by *Roth et al.* (2014a). Such a plume produces a neutral particle density of $\sim5\times10^4$/cm$^3$ at the altitude of the JUICE closest approach (400 km), this value is five times higher than the exospheric density at this height (see Section 6.3). Sources with a lower mass flux will not be distinguishable from the exospheric sources (see Section 6.3). Furthermore, as discussed in the introduction, the event reported in *Roth et al.* (2014a) could be an exceptionally intense event. Rather, much smaller gas releases could be more frequent. Moreover, to show the feasibility of the detection of plume-originating water molecules with confidence, it is better to assume a low mass flux.

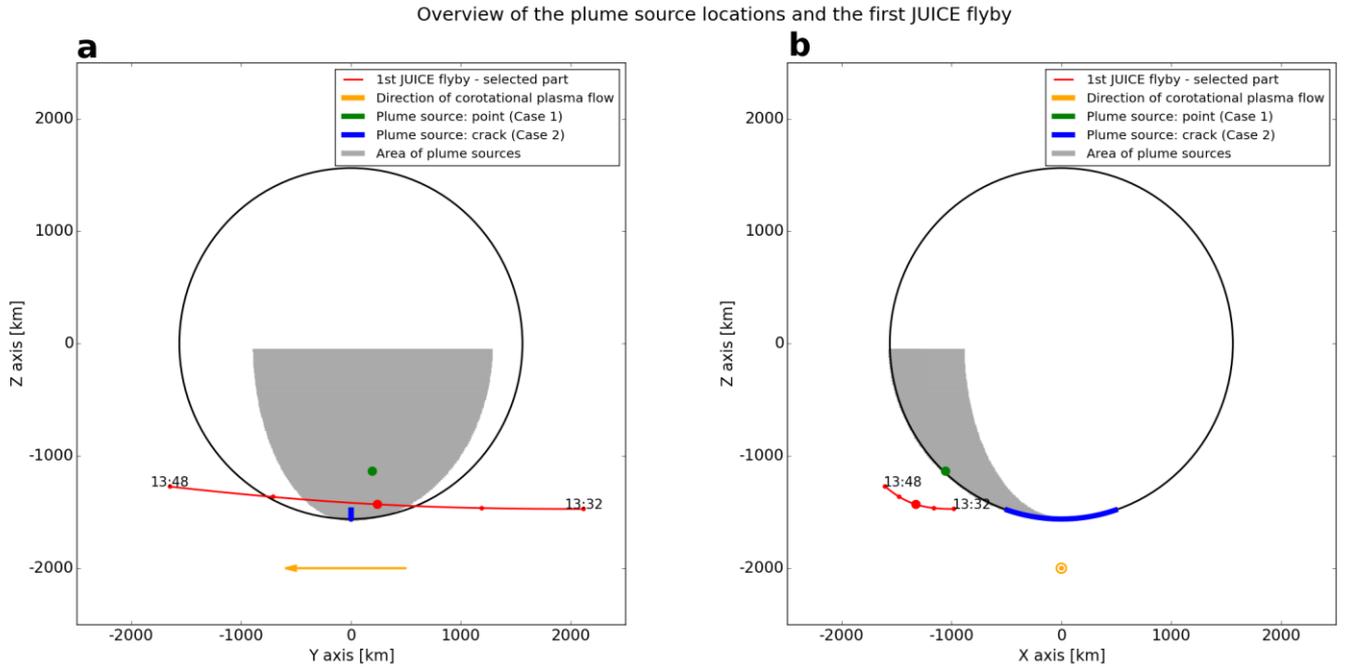

Figure 1: The location of the assumed plume sources together with the geometry of the Europa flyby. (a) *ZY* plane. (b) *XZ* plane. The red line shows the first JUICE flyby. The red dots are separated by four minutes in time. The biggest red dot indicates the closest approach. Indicated with a green dot is the location of a plume source directly below the closest approach. In blue is the location of a 1000 km crack on Europa's south pole. The grey surface indicates the area in which plume sources will be located to investigate how the NIM count rate will vary for different plume locations. The orange arrow shows the direction of Jupiter's corotational plasma.

Each particle has an initial velocity vector that is randomly produced from a Maxwellian velocity distribution corresponding to the 230 K source temperature. A source temperature of 230 K is assumed here based on the estimates made by *Roth et al*. (2014a, supplementary material). The temperature corresponds to a thermal speed of 460 m/s. It should be noted that at a temperature of 230 K water is in fact frozen. A higher value is likely more realistic. However, the results of our simulations do not change significantly for changes in temperature of at least 100 K; therefore we continue using this temperature. Furthermore, a bulk speed perpendicular to the local surface is added. The bulk velocity accounts for the narrow shape a plume escaping from a surface crack presumably has. The bulk speed is assumed to be equal (460 m/s) to the thermal speed. With these initial conditions, the height of the plume will be ~100 km. This is close to the value (200 km) obtained from the observation (*Roth et al.*, 2014a).

With the chosen bulk speed (460 m/s), that is smaller than Europa's escape velocity (2.025 km/s), the average particle will stay well within the Hill sphere of Europa with respect to Jupiter (1.35×$10^4$ km). Thus we neglect the influence of Jupiter's gravity. The average ballistic residence time of a particle with the chosen velocity distribution is less than 1000 seconds. During this time Europa rotates about one degree. Therefore, we also neglect the centrifugal force.

Each launched super particle is traced until the particle impacts on Europa's surface. In this study we neglect any loss processes because of their small contribution. The expected major loss processes of $H_2O$ are photo-reactions, electron impact reactions and ion-neutral interactions (*Johnson et al.*, 2009, *Luchetti et al.*, 2016). The typical lifetime of $H_2O$ due to photo reactions near Europa is $2.24 \times 10^6$ s (*Johnson et al.*, 2009). The lifetime due to electron impact reactions depends on the electron density and electron energy. Electron impact reaction rates are listed in *Johnson* et al. (2009) for electron energies of 20 eV and 250 eV. The typical electron energy at Europa is 100eV (*Kivelson et al.*, 2009), but in this estimation we will assume the 250eV case as a worst case estimate. For the electron density of 110 particles/cm$^3$ (based on *Kivelson et al.*, 2009), the total lifetime due to electron impact reactions at 250eV is $7 \times 10^4$ s. *Luchetti et al.* (2016) also provides an estimate for the lifetime due to ion-neutral charge exchange: $9 \times 10^6$ s. Compared to the typical ballistic resident time of a particle, about 1000 seconds, the lifetime is much longer. During the selected simulation time of 10000 seconds, less than 1.5 % of the total number of particles ($H_2O$) will be lost due to the combined effect of photo reactions, electron impact reactions and ion-neutral charge exchange.

The obtained density profile at each spatial grid cell is then converted to the count rate that the NIM instrument reports. The conversion is linear, represented by 1 particle/cm$^3$ = 5 counts. The intrinsic NIM background noise (Poisson distributed) is 5 counts, for 5-seconds accumulation time. At Europa background noise from penetrating electrons is also expected, which adds ~30 counts, Poisson distributed (*Lasi et al.*, 2016). The combined background counts are estimated to be 35. The plume signal will be obtained by subtracting the background counts from the total (background + plume) counts. To be distinguishable from the instrument and background noise, the (plume) signal to noise ratio should be above one.

We evaluate how the NIM count rate varies along the first JUICE flyby of Europa, for different positions of the plume source. To save computation time we move the trajectory relative to the neutral density distribution obtained from the plume model, rather than running the plume model for each plume position.

Europa's tenuous exosphere also contains $H_2O$ that has been sputtered from the surface ice. Both the sputtered $H_2O$ and $H_2O$ from the plume will contribute to the same $H_2O$ signal that NIM will measure. Models of Europa's exosphere (*Shematovich et al.*, 2005 and *Smyth and Marconi* 2006) show the density of $H_2O$ from 400 km (the altitude of the closest approach) to 1000 km is approximately constant at $10^4$ particles per cm$^3$. The resulting NIM signal is $5 \times 10^4$ counts for 5 seconds integration time. The simulated count rate corresponding to plume-originating particles will be compared to this exospheric signal.

The NIM instrument detects neutral particles with energy lower than ~5eV. It should be noted the particle energy seen by the instrument is the resultant of the actual particle velocity and the spacecraft velocity. The velocity of JUICE relative to Europa is ~4 km/s and the average particle velocity in our simulation is ~0.5 km/s. The combined velocity of 4.5 km/s results in an energy < ~2eV. This is below the upper detection limit; therefore we ignore the effect of the velocity of the spacecraft relative to the neutral particles.

# 4  IONIZED PLUME MODEL

From the simulated spatial profile of the neutral density, we launch ionised super particles from each grid cell. The ion super particles represent many ions travelling together. By calculating the trajectories of the ions under the influence of the Lorentz force, we determine the distribution of ion density and ion flux in each grid cell in a three dimensional space. The directional and energy distribution of the ions at each grid cell are determined as well, which are later used for the ion mass spectrometer detection assessment. Physical and technical simulation parameters are summarised in Table 3 and Table 4 and will be discussed in the following paragraphs.

Table 3: Physical input parameters used for ionised plume particle simulation. Here reference 1 refers to Table 1 in *Johnson et al.* (2009), reference 2 refers to Table 2 in *Kivelson et al.* (2009) and reference 3 refers to Table 3 *Luchetti et al.* (2016).

| Parameter | Value | Reference |
|---|---|---|
| Total ionization rate of $H_2O$ to $H_2O^+$ in the model [$s^{-1}$] | $3.65 \times 10^{-6}$ | 1,3 |
| Electron density [particles/$cm^3$] | 110 | 2 |
| Electron energy [eV] | 100 | 2 |
| Charge of $H_2O^+$ [Coulomb] | $1.602 \times 10^{-19}$ | |
| Magnetic field (B) components [Tesla] | [0, 0, $-415 \times 10^{-9}$] | 2 |
| Bulk velocity of corotating plasma ($\vec{v}_b$) [m/s] | [0, $-76 \times 10^3$, 0] | 2 |
| Electric field E [V/m] | [-0.0315, 0, 0] | $\vec{E} = -\vec{v}_b \times \vec{B}$ |

Table 4: Technical input parameters for the plume ion simulation.

| Parameter | Value |
|---|---|
| Time step length Δt [s] | 0.002 |
| Number of time steps | 30000 |
| Ionization rate per time step [Δt⁻¹] | 5.06×10⁻⁹ |
| Grid cell size [km] | 10×10×10 |

At Europa photoionization and electron impact ionization with $H_2O$ can produce $H_2O^+$, $OH^+$ and $H^+$ (*Johnson et al.*, 2009). Here we focus on $H_2O^+$ ions, created by electron impact ionization and charge exchange. The corresponding ionization rate for electron impact ionization is 2.3×10⁻⁸ cm³ s⁻¹ (*Johnson et al.*, 2009), for 20eV electron energy. *Johnson et al.* (2009) also provides a rate at 250eV of 8×10⁻⁸ cm³ s⁻¹. Here we assume the lower rate, as a conservative estimate (the typical electron energy at Europa is 100eV; *Kivelson et al.*, 2009). For the assumed electron density (110 per cm³; *Kivelson et al.*, 2009) the ionization rate of $H_2O$ to $H_2O^+$ will become 2.53×10⁻⁶/s. The photo-ionization rate of $H_2O$ to $H_2O^+$ is 1.2×10⁻⁸/s. It is significantly smaller than the electron impact ionization rate, thus we ignore photoionization. Recent work suggests the ionization rate of $H_2O$ to $H_2O^+$ due to charge exchange with ionospheric plasma and corotational plasma is about 1.12×10⁻⁶/s (*Luchetti et al.*, 2016). The resulting total ionization rate in this work is 3.65×10⁻⁶/s.

After ionization the particle trajectory will be determined by the Lorentz force that depends on the electric and magnetic fields. We assume homogeneous and perpendicular fields in this study. The magnetic field in our model has a component along the rotation axis of Europa with a magnitude of –415 nT (Table 2 in *Kivelson et al.*, 2009). We neglect the presence of Europa's induced magnetic field (*Khurana et al.*, 1998) for simplicity. The assumed magnetic field configuration is closest to the situation when Europa is crossing the plasma sheet, at which point the perpendicular component of the induced magnetic field to the main field is weakest (~80nT; *Khurana et al.*, 1998).

The electric field considered here is the convection electric field, which is expressed with the following relation: $\vec{E} = -\vec{v}_b \times \vec{B}$. Here $\vec{v}_b$ is the bulk speed of the corotational plasma (76 km/s from Table 2 in *Kivelson et al.*, 2009). In this work the magnetic field is assumed to align with the rotation axis and the corotational flow is along the *y*-axis.

The initial velocity of the neutral particle (~1 km/s) is neglected because it is insignificant compared to the velocity the particle will have as an ion (~100 km/s). The influence of gravity is also ignored because accelerations due to gravity are significantly smaller than the acceleration caused by the Lorentz force. The gravitational acceleration on the surface of Europa is 1.315 m/s². The acceleration by the corotation electric field is $qE/m$ = 1.7×10⁵ m/s², where $q$ is the elementary charge (1.60x10⁻¹⁹ coulomb, and $m$ is the mass of a water molecule 2.98×10⁻²⁶ kg).

We take the JDC instrument as a reference case to assess the detectability of the produced ions. First the distribution of water ions for the viewing directions and energy range of the instrument is simulated. This is converted to differential flux *j(E, Ω)* per energy (*E*) and directional bin (*Ω*), in the unit of particles cm$^{-2}$ sr$^{-1}$ s$^{-1}$ eV$^{-1}$. Subsequently this differential flux is converted into a count rate *r* (in counts per second) using the JDC instrument characteristics (given in Table 1).

$$r(E, \Omega) = j(E, \Omega) G E \Delta t \varepsilon \tag{4-1}$$

Here *G* is the geometric factor, *E* the energy at the middle of the energy bin, *Δt* the measurement time per bin and *ε* accounts for additional efficiencies.

The viewing directions of the instrument are expressed in azimuth and elevation. The azimuth angle is defined as the clockwise angle about the *z*-axis in the *xy*-plane of the reference frame described in Section 3. The azimuth angle ranges from -180° to +180° degree, where 0° azimuth corresponds to the *x*-axis and -90° to the direction from which the corotational plasma is coming. The elevation angle is defined as the angle about the *y*-axis with respect to the *xy*-plane. 0° elevation is parallel to the *xy*-plane and 90° elevation corresponds to a vector parallel and in the direction of the *z*-axis. For the instrument simulations we translate this instrument reference frame to every location for which the JDC simulation is performed, without rotating the frame.

It should be noted the plasma velocity seen by the instrument is the resultant of the actual plasma velocity and the spacecraft velocity. The velocity of JUICE relative to Europa is ~4 km/s. This value is rather small compared to the velocity of the plume ions (~100 km/s). Thus we ignore the effect of the velocity of the spacecraft relative to the plasma.

# 5 RESULTS

## 5.1 NEUTRAL PARTICLES

Figure 2 shows the spatial distribution of the neutral $H_2O$ density from the plume. The plume source in this case is located at the south pole of Europa. Figure 2a shows the spatial density distribution on the scale of Europa, Figure 2b shows a close-up of the source region. In Figure 2a a local elongated increase in density can be observed at the north pole, exactly on the opposite side of the plume source. In this simulation Europa functions as a 'gravity lens'. The gravity lens focusses particles along a focal line instead of on a single focal point.

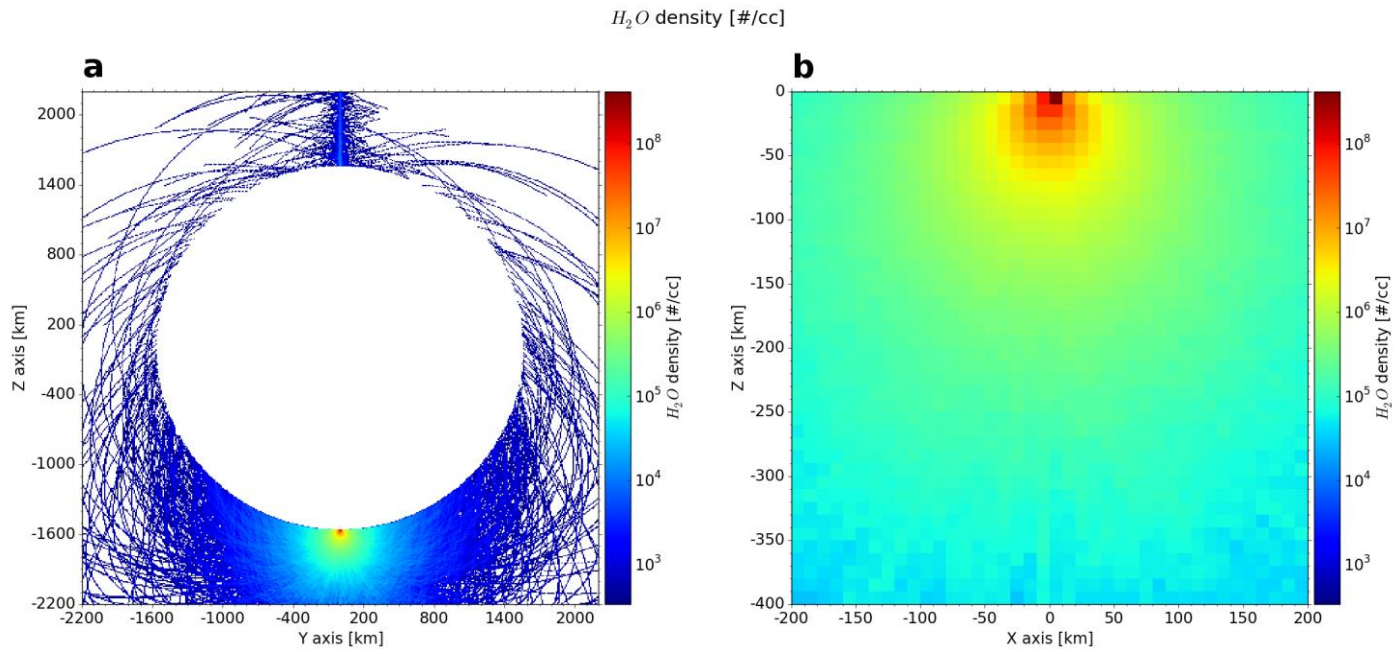

Figure 2: Spatial neutral density distribution ($H_2O$) around Europa resulting from a plume source at the south pole. (a) density distribution on the scale of Europa. (b) south pole close up.

Figure 3a shows how the maximum count rate along the orbit depends on the position of the plume source. The plume positions considered are centred about the projection of the closest approach to the surface of Europa and vary from −45 to +45 degree in latitude and longitude on Europa's surface, the corresponding area is shown in Figure 1. The spacing between the different positions is one degree in latitude and longitude. Here only plume-originating water particles are considered. The expected number of counts (over five seconds integration) is around $2\times10^5$. Throughout the simulation the plume signal is above the standard deviation of the expected NIM noise rate (~6). It should be noted the maximum count rate is not always obtained at the closest approach. In some cases when the plume source is not directly below the closest approach, the maximum count rate is obtained at a different instant in the flyby. Even if the plume source is located 50° away from the projection of the closest approach on the surface, the NIM instrument can still detect the neutral particles with $10^3$ counts. Figure 3b shows the simulated NIM count rate along the trajectory of the first JUICE flyby, resulting from the combination of the ambient exospheric water and the plume-originating water. A constant count rate of $5\times10^4$/5s due to exospheric water is assumed (*Shematovich et al.*, 2005 and *Smyth and Marconi* 2006). Furthermore we assume the plume is located directly below the closest approach of the JUICE spacecraft to Europa (Figure 1). The presence of this small plume creates a bump of about one order of magnitude in the total water signal.

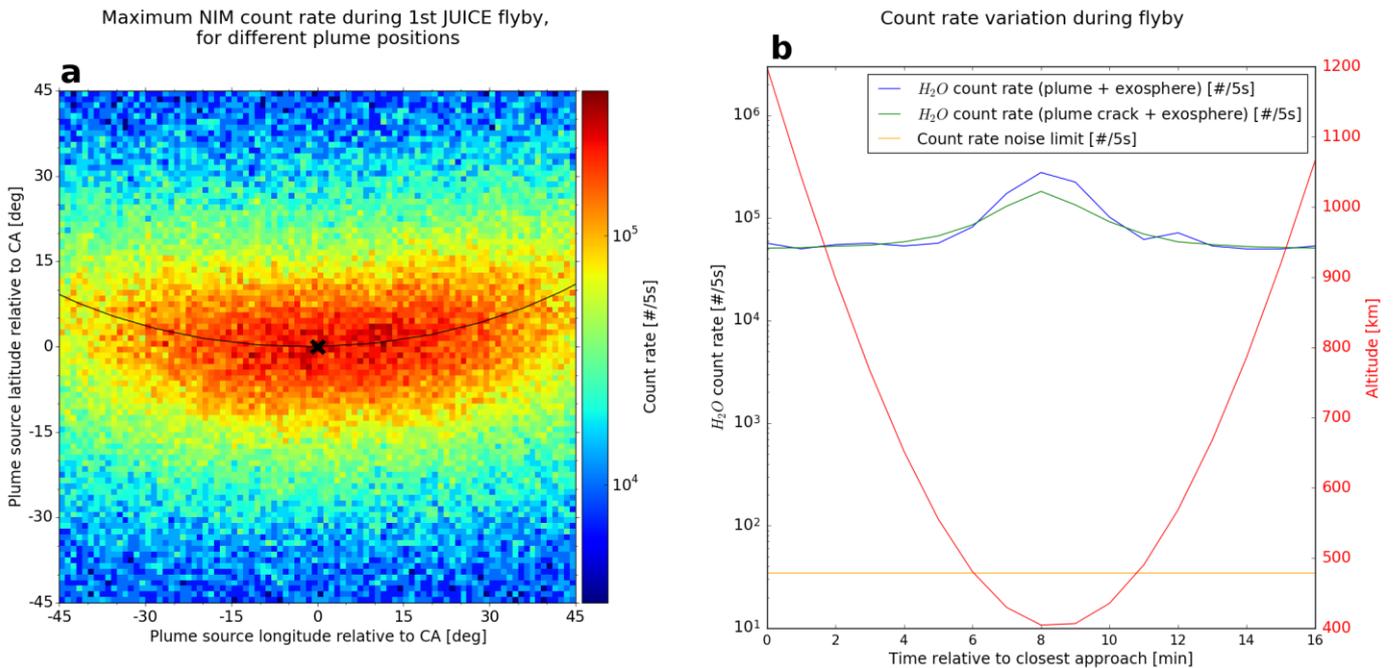

Figure 3: (a) NIM maximum $H_2O$ count rate (plume only) along the JUICE flyby trajectory for different positions of the plume. The black line shows the projection of the spacecraft trajectory on the surface of Europa, the cross mark indicates the closest approach. The corresponding area is shown in Figure 1. (b) In blue: NIM $H_2O$ count rate (plume + exospheric) during the first JUICE flyby when the plume (point source) is directly below the closest approach of JUICE to Europa (Figure 1). The time of the closest approach, as currently planned, is 2031-02-13 13:40:00 UT. In green: NIM $H_2O$ count rate when the plume source is a 1000 km crack. The crack is located along the meridian passing through the projection of the closest approach on the surface of Europa (Figure 1) and is centred on the latitude of the projection of the closest approach.

The spatial density distribution resulting from the 1000 km crack, which is about perpendicular to the flyby trajectory, is shown in Figure 4a. The crack position is shown in Figure 1. By comparing Figure 2 and Figure 4a it can be seen that the peak density close to the source is about two orders of magnitude smaller than in the case of a single point source. At the altitude of the closest approach (400 km) the density varies between $10^4$ and $10^5$ particles per $cm^3$. This value is similar to that of the single point source (Figure 2) at 400 km altitude. Then we evaluate how the maximum count rate along the orbit changes corresponding to changes in the position of the centre point of the crack. The result is shown in Figure 4b. By comparing to Figure 3a it can be concluded that the maximum count rate is lowered for plume positions closer than ±15 degree latitude and ±30 degree longitude. Outside this area the count rate is very similar to the single point source case (Figure 3a).

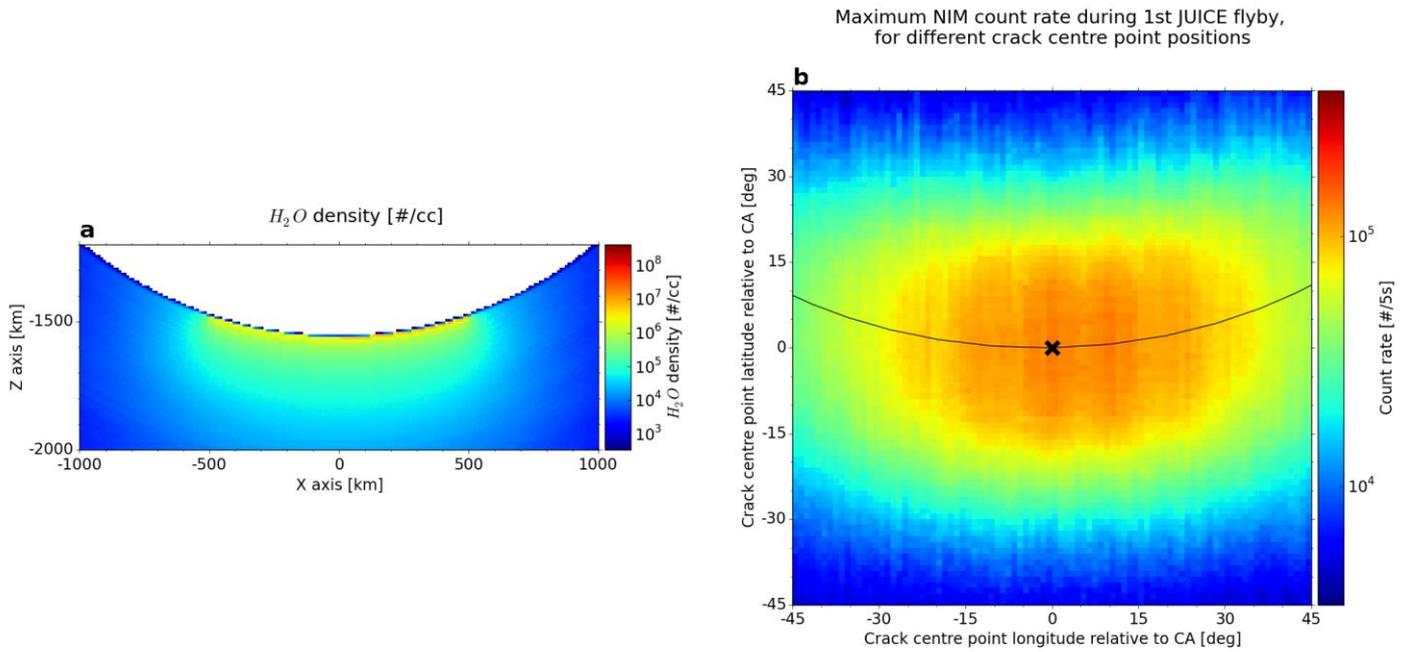

Figure 4: (a) Spatial neutral density distribution ($H_2O$) around Europa resulting from a plume source crack at the south pole. (b) NIM maximum count rate along the JUICE flyby for different positions of the crack centre point. The black line shows the projection of the spacecraft trajectory on the surface of Europa, the cross mark indicates the closest approach.

Figure 3b shows the simulated NIM count rate along the trajectory of the first JUICE flyby, resulting from the combination of the ambient exospheric water and the plume-originating water. Here it is assumed that the plume source is a crack of 1000 km. The plume source is located along the meridian passing through the projection of the closest approach on the surface of Europa (Figure 1) and is centred on the latitude of the projection of the closest approach. A constant count rate of $5\times10^4$/5s due to exospheric water is assumed (*Shematovich et al.*, 2005 and *Smyth*). Compared with the case of a single point source the bump in the count rate caused by the plume is still recognizable, albeit lower. The difference in the count rate is less than one order of magnitude, 66% at the maximum.

## 5.2 IONIZED PARTICLES

Figure 5a shows the spatial density distribution of $H_2O^+$ ions in a plane that is located 50 km below (in the negative *z*-direction) from the plume source (and thus the south pole of Europa); note that the plane does not intersect with Europa. The ions are produced by a plume, of which the (point) source is located at Europa's south pole (Figure 2). Figure 5b shows the corresponding spatial distribution of the flux of $H_2O^+$. The flux is obtained by multiplying the ion density in each grid cell with the average speed of the super particles in each cell. The point in the centre of Figure 5a and Figure 5b is closest to the plume source, therefore the density of the neutral particles is highest there. Thus, more ions will originate from this point than areas further away from the source. By considering that all plume ions are transported along the *y*-axis, it can be explained why we can recognize the signature of the cycloid trajectories of the particles from close to the plume source in the ion density and flux. Areas with high flux are of most interest for ion observations with JDC, since the count rate will be highest there.

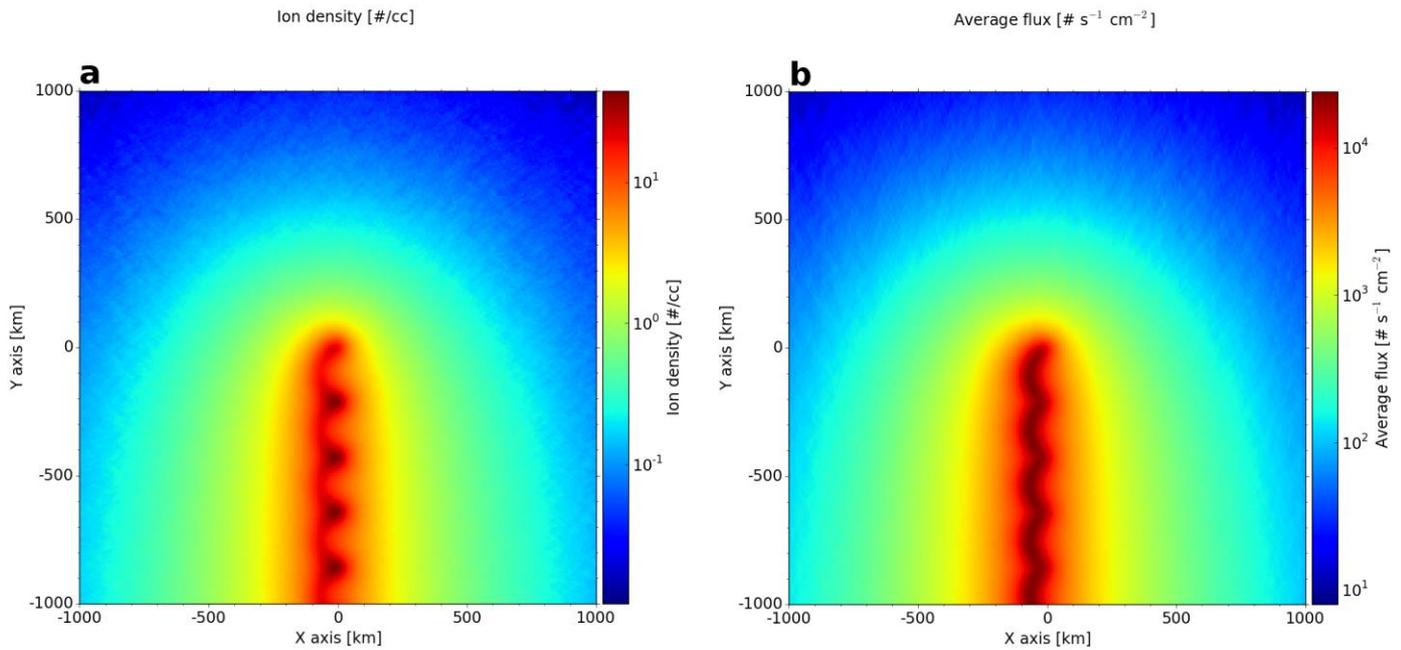

Figure 5: ion density(a) and ion flux (b) in a plane 50 km below (in negative z-direction) Europa's south pole. Both (a) and (b) are obtained from a single-point source neutral plume model (Figure 2).

Figure 6 shows the count rates expected to be observed by the JDC instrument. A plume has been located directly below the closest approach of the first JUICE flyby for this simulation (Figure 1). Figure 6a shows the count rate per energy and azimuth at the closest approach of the flyby. The elevation angle of the instrument is zero, so it is looking in the *xy* plane, namely, the same plane as Figure 5. Figure 6b shows how the count rate varies over time during the flyby for the looking direction from Figure 6a with the highest count rate. This is the -90° azimuth angle, which is looking in the same direction as the negative *y*-axis (Figure 5) and thus the direction from which the corotational plasma is coming. The plume ion signal is ~3×10$^3$ counts at the maximum (assuming one second integration time). This is bigger than the expected JDC noise signal (~ 100 counts for one second integration time). Figure 6b also shows how the count rate varies over time for the optimal looking direction (in terms of count rate). It can also be seen that the count rate is asymmetrical along the trajectory.

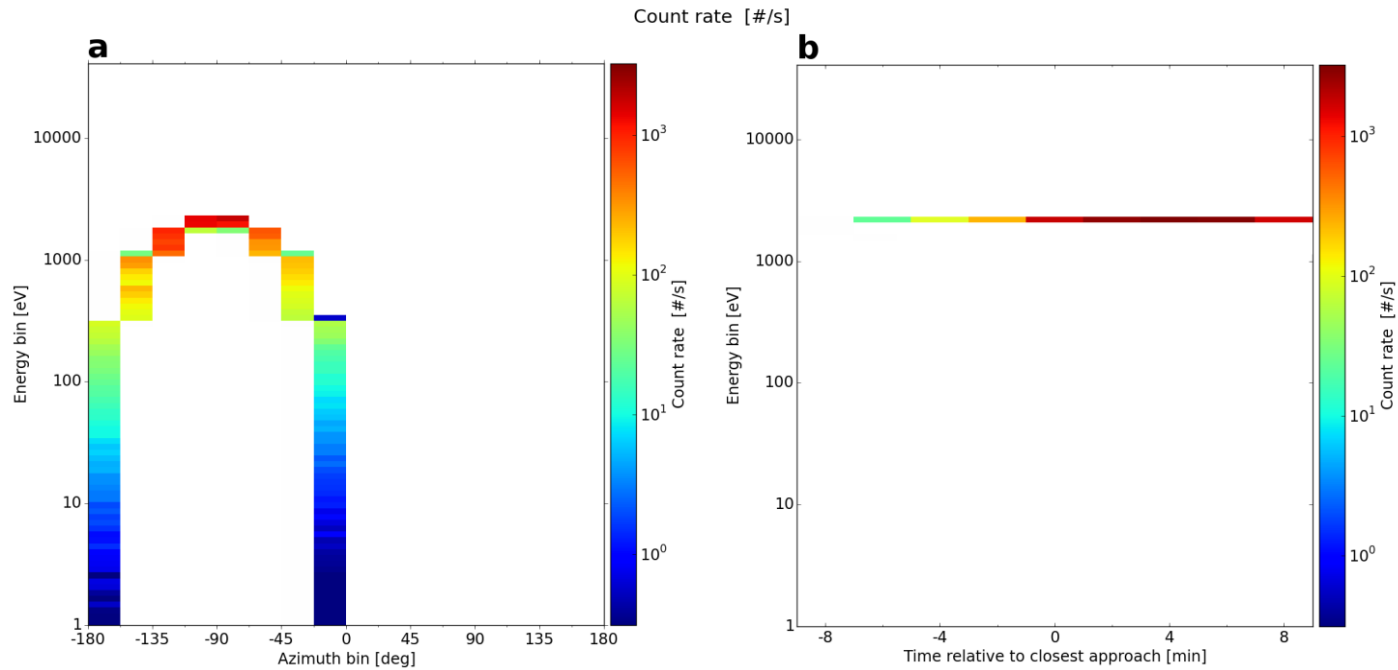

Figure 6: (a) Count rate ($H_2O^+$) per direction and energy bin. The simulation corresponds to the closest approach of the first JUICE flyby (at 13:40). The elevation angle is zero. The plume source is located directly below the closest approach (Figure 1). (b) Count rate ($H_2O^+$) per energy bin versus time, for the azimuth bin -90° to –67.5° from (a), corresponding to the direction from which the corotational plasma flow is coming. The time of the closest approach, as currently planned, is 2031-02-13 13:40:00 UT.

In Figure 7 the $H_2O^+$ count rate in the optimal viewing direction (corresponding to the maximum count rate) is shown for three different cases. The first case shows the count rate if there is only an $H_2O$ exosphere present, thus no plume. The exospheric density profile is obtained from *Shematovich et al.* (2005) (Figure 1 therein). This model is chosen over the one in *Smyth and Marconi* (2006) since the maximum $H_2O$ density, and therefore the resulting maximum $H_2O^+$ density, is higher in *Shematovich et al.* (2005). Thus *Shematovich et al.* (2005) represents a worst case scenario. The second case shows the total count rate if the plume (Figure 6b) and the exosphere are present at the same time. Finally, the third case shows how the second case changes when the plume source is not a single point source, but a source crack of 1000 km. The crack differs from the one shown in Figure 1 in the sense that it is oriented along the meridian of the projection of the closest approach on the surface and centred on the latitude of the projection of the closest approach. At the maximum the presence of the plume point source creates an increase in the $H_2O^+$ count rate of a factor ~2 and the plume crack an increase of a factor ~1.75.

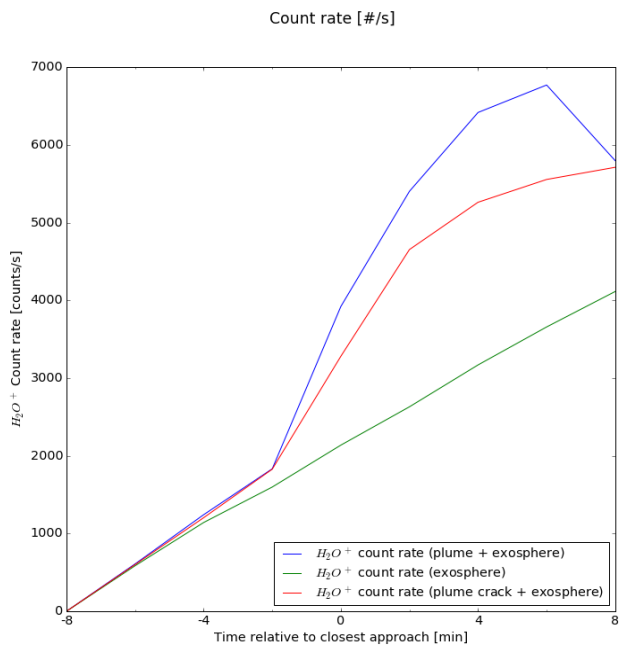

Figure 7: JDC $H_2O^+$ count rate during the first JUICE flyby for three different cases. Blue line: count rate if there is an $H_2O$ exosphere and a plume. The plume (point source) is directly below the closest approach of JUICE to Europa (Figure 1). Red line: count rate if there is an $H_2O$ exosphere and plume. The plume source is a 1000 km crack along the meridian passing through the projection of the closest approach on the surface of Europa (Figure 1). The crack is centred on the latitude of the projection of the closest approach. Green line: count rate if there is only an $H_2O$ exosphere.

# 6 DISCUSSION

## 6.1 DETECTABILITY OF THE PLUME PARTICLES

It can be seen in Figure 3 that the signal of neutral plume particles is well above the noise level of the NIM instrument (Table 1) for the wide range of plume positions considered for this plot. In Figure 3a it can be seen that the count rate is around $2 \times 10^5$/5s. Assuming five seconds integration time and that the background is about 35 counts (Table 1), the signal to noise ratio is ~5700 at this point, therefore also allowing the detection of trace chemical species contained in the water plume. Thus, we conclude that the NIM instrument can detect the $H_2O$ molecules from the plume.

From Figure 6 it can be concluded that the signal from ionized plume particles is distinguishable from the background noise due to penetrating radiation in the JDC instrument. The plume ion signal is ~3×10³ counts at the maximum, assuming one second integration time. The background is around 100 counts, also for or one second accumulation time (Table 1). We assume the background signal is white noise. This results in a signal-to-noise ratio of ~33. From Figure 6b it can be concluded that the count rate is asymmetric with respect to the closest approach; a higher count rate is seen after the spacecraft has passed the point of closest approach. The spacecraft is moving in the same direction as the corotation plasma flow and thus the direction in which the plasma ions are transported (as can be seen in Figure 5).

The simulation results in this work are all performed for the first JUICE flyby of Europa. The second planned flyby is similar, but will happen at higher latitudes. It will approach Europa at a similar closest distance (~400 km) and will also fly along the direction of the corotation plasma. In other words, the same conclusions can be formulated for the second flyby.

### 6.2 Difference between a single source and a crack source

In the case of a 1000 km crack model, the spatial density distribution of $H_2O$ close to the source can be more than one order of magnitude lower than in the case of a single point source model (Figure 4a). This will negatively affect the detectability due to a lower signal-to-noise ratio. However, at the altitude of the JUICE closest approach (~400 km), the count rate is less than one order of magnitude lower (Figure 3b and Figure 4b). The difference is limited to positions of the crack centre point ±15 degree latitude and ±30 degree longitude from the closest approach. Outside of this area the count rate is comparable. At ±20 degree latitude a small increase can even be observed. Regarding the detection of $H_2O^+$, Figure 7 shows that the difference in the $H_2O^+$ count rate between the crack source and the single source is at most ~35% for the optimal viewing direction. Thus, the source model will not impact the $H_2O$ or $H_2O^+$ detection feasibility significantly.

### 6.3 Separation with exospheric signal

In the count rate corresponding to water particles (Figure 3b) the presence of the plume can be recognized as a local increase of about one order of magnitude. This suggests that the signature of $H_2O$ originating from small plumes (mass flux 1 kg/s) can be recognized in the $H_2O$ signal measured by NIM. However, the modelled exospheric density distribution should be confirmed with observations. This is even more important for plumes that are not positioned directly below the closest approach and for which the signature observed along the flyby is weaker. As currently planned, JUICE will make only two flybys of Europa. This offers the possibility for only a limited number of in situ observations to understand the density distribution in Europa's exosphere. Complementary observations of the exospheric density distribution by the remote sensing instruments on JUICE will be required to obtain the full exospheric particle population.

In Figure 3a it can be seen that the plume particle count rate does not drop significantly for source positions that are 30 degrees in longitude and 10 degree in latitude away from the closest approach. It seems plausible that these plumes will be detectable as well. The position of the furthest removed plume that can still be separated from the ambient water particles depends on the knowledge of the exospheric density distribution.

Models of Europa's exosphere (*Shematovich et al.*, 2005; *Smyth and Marconi* 2006; *Wurz et al.* 2014) show the density of $H_2O$ peaks from about $10^4$ to $10^6$ particles per $cm^3$ close to the surface, and at the altitude of the closest approach of the JUICE flyby (400 km) it drops to $10^4/cm^3$. In our model, the plume-originated density close to the surface is $\sim 10^8/cm^3$, and a number density of $5 \times 10^4/cm^3$ is expected at the altitude of the closest approach (Figure 2b). The exospheric neutral particles are ionized in the same way as the plume ions and will conduct cycloid trajectories as well. Because of this the exospheric ions and the plume ions have the same energy and directional distribution. Therefore they cannot be directly separated. Distinguishing the plume and exospheric signals can be aided by the difference in spatial distribution they have (Figure 7). Plume ions form a collimated beam of ions (an 'ion plume'), while the exospheric ions are more globally distributed. Secondly because the neutral plume particle density is bigger, the signal of the plume-originating ions will be bigger than that of the exospheric ions (FIGURE 7). However, the distribution of the ambient $H_2O^+$ density from observations is needed. A back-tracking method can be used to trace particles back in time and determine their point of origin, which in the case of a plume would be a confined region. This method has been employed before at the Moon, to trace back protons to the dayside of the Moon (*Futaana et al.*, 2003), and in the vicinity of Mars to distinguish Phobos-originating protons (*Futaana et al.*, 2010).

## 6.4 SEPARATION FROM COROTATIONAL PLASMA

Near Europa $O^+$ and $O^{++}$ are the dominant ions (*Johnson et al.*, 2009; *Rubin et al.*, 2015). JDC does not have sufficient mass resolution (M/ΔM up to 30) to distinguish between $O^+$ or $O^{++}$ and $H_2O^+$. However there is a difference in energy and direction distribution between $H_2O^+$ and $O^+$ or $O^{++}$. The plume ions propagate along cycloid trajectories, along these trajectories their direction and energy changes, resulting in the direction and energy distribution shown in Figure 6a. The corotational plasma particles are not following cycloid trajectories and will not exhibit the same variation in direction and energy. They will rather be detected as particles coming from a single energy and direction that will coincide with the bins with the maximum count rate in Figure 6a. Thus, the difference in distribution can be used to distinghuish $H_2O^+$ from the oxygen ions in the corotational plasma.

## 6.5 Practical implications for ionized plume particle measurements

Not all the azimuth, energy and elevation bins of the JDC instrument can be sampled simultaneously. Therefore, it should be investigated if one second integration time per bin is realistic to accumulate the counts up a significant level. The JDC instrument has 12 elevation bins and 16 azimuth bins. The number of energy bins can be changed, where 96 bins are the maximum. Sampling one energy step takes 7.8125 ms. The azimuth bins are sampled simultaneously. Figure 6a and Figure 6b correspond to one elevation bin. Namely the elevation bin that contains the plane that is parallel to the **E × B** direction. If only this elevation bin would be sampled, 96 seconds would be needed to provide one second of integration time per bin and accumulate the counts up to the level discussed in paragraphs 6.1 and 6.3. Within 96 seconds JUICE will travel approximately 384 km. Plasma conditions over such a distance can change strongly, in particular when the spacecraft passes the trajectories of the particles originating from close to the plume source (Figure 5). To decrease the time of measurement different strategies can be employed. The number of elevation bins that are scanned can be limited, since the plume particles are propagating in the **E × B** direction. Differently the integration time could be lowered. If it is lowered to 0.1 seconds per bin, the plume signal strength would be ~900 counts, which is still a significant number of counts. Alternatively the bins could be collapsed. Furthermore, the JUICE spacecraft, as presently planned, will fly along the direction of the corotational flow. This is beneficial for the measurements, since the plume ions are propagating in the same direction (see Figure 2a and Figure 5). Therefore the conditions of the ion plume will not change as much as when the ion plume would be crossed perpendicularly. This allows the possibility to increase the measurement time.

## 6.6 Model limitations

In the model, the ion trajectories are solely determined by **E × B** drift caused by a perpendicular electric and magnetic field. The electric field is calculated as follows: $\vec{E} = -\vec{v}_b \times \vec{B}$. In which $\vec{v}_b$ is assumed to be the bulk velocity of the corotation plasma. However, this is a simplification. In principle, $\vec{v}_b$ is the average of the bulk velocity of the corotation plasma flow ($\vec{v}_{b\ cp}$) and the plume-originating ions ($\vec{v}_{b\ plume}$):

$$\vec{v}_b = \frac{n_{cp}\vec{v}_{b\ cp} + n_{plume}\vec{v}_{b\ plume}}{n_{cp} + n_{plume}}$$

When the density of the corotation plasma ($n_{cp}$) is higher than that of the plume ions ($n_{plume}$), $\vec{v}_b$ will be dominated by the bulk velocity of the corotation plasma. Then the electric field can be approximated as: $\vec{E} = -\vec{v}_{b\ cp} \times \vec{B}$. For a single plume source with a mass flux of 1 kg/s the plume ion density can exceed 10 particles per cm³ in the regions where the ions originate from near the source (see Figure 5a). This is rather close to the ion density at Europa, ranging from 12 to 170 particles per cm³ (from *Kivelson et al.*, 2009). However, in most of the simulation space the density does not exceed 10 particles per cm³. Using the equation above and assuming that $n_{cp}$ is 100 particles per cm³, $\vec{v}_{b\ cp}$ is 76 km/s (Table 3), $n_{plume}$ is 10 particles per cm³ and $\vec{v}_{b\ plume}$ is 1 km/s; gives a bulk velocity of 69 km/s. This value is not significantly different from $\vec{v}_{b\ cp}$. Therefore we consider the assumption $\vec{E} = -\vec{v}_{b\ cp} \times \vec{B}$ acceptable.

The assumption is no longer satisfied if the density of the plume ions becomes large compared to that of the ambient ions in the corotation plasma flow (~100 km/s). For example in the case reported in *Roth et al., 2*014a. The bulk velocity of the plasma will be dominated by the velocity of the plume-originating ions (~1 km/s). Significant slow down of the corotation plasma flow associated with plume activity has indeed been observed at Enceladus (*Tokar et al.*, 2006). The assumed convectional electric field cannot be calculated from the corotation flow anymore. The electric field and the resulting ion trajectories will become significantly different. Thus, the mass loading of the flow of corotating plasma will change the flow and distribution of plasma near Europa (*Rubin et al.*, 2015). Such a situation cannot be handled by the model that we used. Instead, a different modelling approach is needed, such as hybrid or magnetohydrodynamic models. Furthermore, when the flow of the ions is significantly slowed down the relative velocity of the spacecraft becomes a factor of importance.

The high-flux plumes also influence the ionisation. For an electron temperature of 100 eV (*Kivelson et al.*, 2009) the mean free path of an electron is ~150 km, for a total $H_2O$ – electron scattering cross sectional area of 7×10$^{-16}$ cm² (from *Itakawa and Manson* 2005) and an $H_2O$ density of ~10⁸ particles per cm³ (from Figure 2a). If the neutral density would be several orders of magnitude larger, the plume will not be ionised equally everywhere, in particular an area close to the source will be free of plume ions.

In our model, we neglect the neutral-neutral collisions. Close to the source, the plume particles may experience multiple collisions, behaving more like a fluid. A collisional model should be used in the vicinity of the plume source. On the other hand, in *Hurley et al.* (2015) an Enceladus plume model was discussed that uses the same approach for launching neutral particles as used in this work. *Hurley et al*. (2015) stated that a velocity distribution that is a combination of a thermal velocity and a bulk speed (referred to as a drifting Maxwellian) is appropriate outside the collisional region of the plume. This collisional region is only a few kilometres high, in the case of Enceladus' plumes (*Yeoh et al*., 2015). Although the escape velocity is different between Enceladus and Europa (0.239 km/s and 2.025 km/s, respectively), the initial velocities (900 m/s and 450 m/s) and the mass fluxes (0.2 kg/s and 1 kg/s, respectively) in the Enceladus plume model in *Yeoh et al.* (2015) are similar to the plume properties we assume. With a velocity of 0.5 km/s it takes a particle about 20 seconds to reach 10 km altitude. Considering that Europa's gravitational acceleration is 1.314 m/s$^2$, gravity will decelerate the particle ~25 m/s within this time. This is small compared to the initial velocity, thus in the first few kilometres above the surface gravity has not significantly altered the trajectories yet. Therefore we expect the difference in escape velocity to be unimportant. This suggests that the size of the collisional region of the 1 kg/s Europa plume we assume is comparable to the Enceladus case described just before. Therefore, at large distances from the collisional region, such as that of the JUICE flyby (> 400 km), the plume model we assume is applicable. Furthermore, by comparing analytical solutions for collision-less flow and direct simulation Monte Carlo (DSMC) model results, *Cai and Wang* (2012) and *Cai et al.* (2013) suggested that collision-less flow solutions can still be used to obtain a first order approximation of a collisional supersonic flow. They explained the models agree because the particles travelling downstream with a high bulk speed do not have the time to collide and propagate in the direction perpendicular to the bulk speed.

In this work only $H_2O^+$ as an ionized product for $H_2O$ is considered. However, Johnson et al. (2009) also lists ionization rates of $H_2O$ that give $OH^+$ and $H^+$. The electron impact reactions that produce $OH^+$ and $H^+$ have the same order of magnitude ($10^{-6} - 10^{-7}$/s) as the rate that produces $H_2O^+$. Thus these species also can be detected by JDC.

# 7 CONCLUSION

We simulated the detection of $H_2O$ and $H_2O^+$, by respectively NIM and JDC. The signal-to-noise ratios in the optimal cases, in terms of plume position with respect to JUICE, are ~5700 for NIM and ~33 for JDC, if only a background due to penetrating radiation is assumed. The results show that the particles of the low mass flux plumes (1 kg/s) can be detected with large margins. By comparing the density distributions from a point source and a 1000 km crack, we conclude that the difference does not significantly influence the density distributions and detectability.

Furthermore, we investigated the distinction of plume-originating $H_2O$ and exospheric $H_2O$. The results suggest the plume signal is recognisable as a temporal signal enhancement of one order of magnitude in the $H_2O$ count rate during 6 minutes. However, the density distribution obtained from models has still to be validated by observations. For the separation of $H_2O^+$ from ambient $H_2O^+$, the difference in spatial distribution can be important. However, knowledge of the ambient $H_2O^+$ density is needed for this as well. Additionally a particle back-tracing model is suggested as a data analysis tool to separate the signals.

Some measures could be taken to increase the number of detected counts with JDC. Sectors can be collapsed or neglected, and/or the integration time can be reduced. It is beneficial for the JDC observations that the flyby is along the direction of the corotational plasma (and the plume ions) because the signal of plume ions will persist longer and therefore a longer integration time can be achieved.

Several recommendations for future Europa plume particle models can be formulated. Future models that require more accurate knowledge of the neutral particle distribution should consider particle collisions. Also the effect of ionized plume models on the electric field should be taken into account. To investigate plume observations in more realistic magnetic field configurations it is also needed to include the effect of Europa's induced magnetic field.

## Acknowledgements
The contribution of Karl-Heinz Glassmeier is financially supported by the German Bundesministerium für Wirtschaft und Energie and the Deutsches Zentrum für Luft und Raumfahrt under grant 50QJ1501.